\documentclass[11pt,twocolumn]{article}
\usepackage[authoryear]{natbib}
\usepackage{graphicx}

\begin{document}

\title{Large-scale asymmetry between clockwise and counterclockwise galaxies revisited}

\author{Lior Shamir \\ Kansas State University \\ email: lshamir@mtu.edu}




\date{}

\maketitle

\abstract{
The ability of digital sky surveys to collect and store very large amounts of data provides completely new ways to study the local universe. Perhaps one of the most provocative observations reported with such tools is the asymmetry between galaxies with clockwise and counterclockwise spin patterns. Here I use $\sim1.7\cdot10^5$ spiral galaxies from SDSS and sort them by their spin patterns (clockwise or counterclockwise) to identify and profile a possible large-scale pattern of the distribution of galaxy spin patterns as observed from Earth. The analysis shows asymmetry between the number of clockwise and counterclockwise spiral galaxies imaged by SDSS, and a dipole axis. These findings largely agree with previous reports using smaller datasets. The probability of the differences between the number of galaxies to occur by chance is $(P<4\cdot10^{-9})$, and the probability of an asymmetry axis to occur by mere chance is $(P<1.4\cdot10^{-5})$. 
}


\section{Introduction}
\label{introduction}

Modern astronomical digital sky surveys are enabled by robotic telescopes collecting massive databases of astronomical data. The availability of these databases enable observations of the local universe by analyzing a very large number of astronomical objects, an approach that was not possible in the pre-information era. One of the most provocative observations that have been noted using these databases is the suspected large-scale patterns of galaxy rotation direction. Previous experiments showed non-random patterns of the distribution of clockwise and counterclockwise galaxies \citep{longo2011detection,shamir2012handedness,shamir2013color,hoehn2014characteristics,shamir2016asymmetry,shamir2017colour,shamir2017photometric,shamir2017large,lee2019galaxy,lee2019mysterious}.

The asymmetry has been shown with manually annotated galaxies \cite{longo2011detection}, as well as with automatically annotated galaxies \citep{shamir2012handedness}, both showing differences in the number of clockwise and counterclockwise galaxies as observed from Earth by SDSS, and exhibiting a dipole axis \citep{longo2011detection,shamir2012handedness}. Other observations showed photometric asymmetry between clockwise and counterclockwise galaxies \citep{shamir2013color}. The marginal statistical significance of the difference \citep{hoehn2014characteristics} was improved by the use of machine learning to show that the photometric variables of a galaxy can be used to predict its spin pattern with accuracy much higher than mere chance \citep{shamir2016asymmetry}. More recent work using a larger set of galaxies showed clear and statistically significant photometric differences between clockwise and counterclockwise galaxies in SDSS \citep{shamir2017colour,shamir2017photometric} as well as Pan-STARRS galaxies \citep{shamir2017large}, showing that the two digital sky surveys identify similar asymmetry between the photometry of clockwise and counterclockwise galaxies \citep{shamir2017large}. Studies using a smaller number of 445 galaxies showed that spin direction of neighboring galaxies can correlate even when the galaxies are too far to have any gravitational interaction \citep{lee2019mysterious}. Some evidence also showed alignment between the polarization of quasars and the large-scale structure \citep{hutsemekers2014alignment}.

A pre-information era experiment showed no difference between clockwise and counterclockwise galaxies \citep{iye1991catalog}. However, as no digital sky surveys were available at the time, the dataset contained just a few thousand galaxies, which is insufficient to show a statistical significance of the asymmetry. Another attempt was made by using crowdsourcing analysis of SDSS galaxies \citep{land2008galaxy}, which also showed no statistically significant difference between clockwise and counterclockwise galaxies.  However, that study also showed that untrained volunteers do not excel in the task of classifying galaxies by their spin patterns, leading to an unclean dataset, heavily biased by the human perception \citep{land2008galaxy}. When comparing the photometry of just the annotations on which 95\% or more of the volunteers agreed on, the photometric differences between clockwise and counterclockwise galaxies was aligned with the same photometric asymmetry observed in \citep{shamir2017photometric}, but the selection of the galaxies makes the dataset too small to be considered statistically significance \citep{shamir2017large}.

Here I revisit the comparison of the number of clockwise and counterclockwise galaxies by using $\sim1.7\cdot10^5$ SDSS galaxies annotated automatically by their spin patterns. The paper follows the experiments in \citep{longo2011detection,shamir2012handedness}, but with more and cleaner data.

\section{Data}
\label{sec:data}

The data used in this study was taken from the Sloan Digital Sky Survey. The initial dataset was a catalog of $\sim3\cdot10^6$ SDSS galaxies with i magnitude smaller than 18 and Petrosian radius larger than 5.5'' \citep{kuminski2016computer}. That selection ensured that the galaxies are sufficiently large and sufficiently bright to identify their morphology, as the vast majority of SDSS galaxies are too small and faint for a reliable analysis of their shape.  

The galaxies in the catalog were assigned with the certainty of their broad morphological classification of elliptical and spiral galaxies, such that a certainty value close to 0.5 indicates that the certainty of the classification is low, while a certainty value close to 1 indicates that the classification of the galaxy is most likely correct. A detailed description of the catalog is available in \citep{kuminski2016computer}. Experiments with the automatic annotations of 45,377 included in the catalog and also classified by {\it Galaxy Zoo} as debiased ``superclean'' showed that galaxies that were classified as spiral galaxies with certainty higher than 0.54 were in $\sim$98\% of the cases in agreement with the debiased ``superclean'' Galaxy Zoo annotations \citep{kuminski2016computer}, and therefore it is reasonable to assume that galaxies classified as spiral in certainty higher than 0.54 are indeed spiral galaxies. With 0.54 as threshold, the dataset included 740,908 galaxies annotated automatically as spiral \citep{kuminski2016computer}. It should be noted that Galaxy Zoo data was used to assess the accuracy of the catalog in comparison to manual annotation, but was not used for any classification of the galaxies, which was all done in a fully automatic manner.

\begin{figure*}
\includegraphics[scale=1.0]{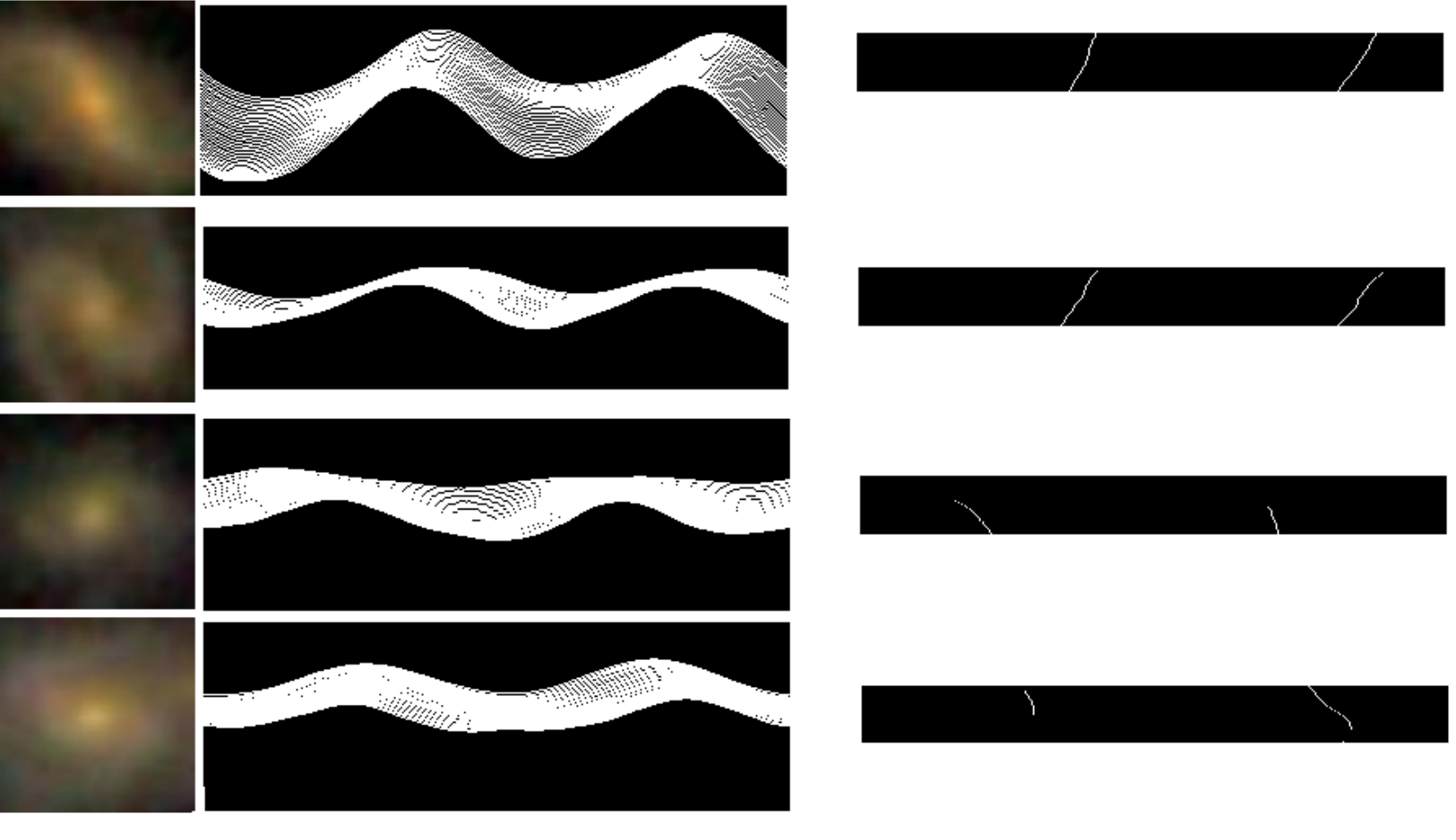}
\caption{Original images (left), their corresponding radial intensity plots (middle), and the peaks detected in the radial intensity plots (right). The sign of the lines of the peaks determines the curve of the arms of the galaxy, and therefore also the direction of rotation.}
\label{redial_intensity_plots}
\end{figure*}

The set of spiral galaxies was then separated into clockwise and counterclockwise galaxies using the Ganalyzer tool \citep{shamir2011ganalyzer,ganalyzer_ascl} as was done in \citep{shamir2012handedness,hoehn2014characteristics,shamir2016asymmetry,shamir2017colour,shamir2017photometric,shamir2017large}. Ganalyzer transforms the galaxy images into their radial intensity plots, which are images of dimensionality of 360$\times$35, such that the X axis is the polar angle (in degrees) and the Y axis is the radial distance in percents of the galaxy radius. That is, the value of the pixel $(x,y)$ in the radial intensity plot is the median value of the 5$\times$5 pixels around $(O_x+\sin(\theta) \cdot r,O_y-\cos(\theta)\cdot r)$ in the original image, where $\theta$ is the polar angle and {\it r} is the radial distance. Ganalyzer then applies peak detection to identify groups of peaks along the horizontal lines of the radial intensity plot \citep{shamir2011ganalyzer}. Figure~\ref{redial_intensity_plots} shows examples of original galaxy images, the transforms into radial intensity plots, and the peaks detected in the radial intensity plots.

Since arm pixels are expected to be brighter than non-arm pixels at the same distance from the galaxy center, the groups of peaks identify the galaxy arms. The vertical lines of the peaks detected in the radial intensity plot correspond to the curve of the arm, and linear regression is applied to measure that curve. The sign of the curve indicates whether the direction of the arm is clockwise or counterclockwise. The algorithm is described thoroughly with examples and a detailed performance analysis in \citep{shamir2011ganalyzer}, as well as in \citep{hoehn2014characteristics,shamir2012handedness}, and its application to the galaxy dataset used in this study is described in \citep{shamir2017colour,shamir2017photometric,shamir2017large}. 

Because many galaxies identified as spiral do not have a clear spin pattern or can be missclassified by the algorithm, only galaxies which their spin pattern was classified with high certainty were used. To avoid galaxies with unclear spin pattern classification, only galaxies that had linear regression with at least 10 points (10 peaks in the radial intensity plot) were used, and all galaxies that did not have at least 10 peaks were ignored. Also, at least 75\% of the peaks were expected to be aligned in one direction (clockwise or counterclockwise), and galaxies that did not meet that criterion were also ignored. Manual inspection of 200 randomly selected clockwise galaxies and 200 randomly selected counterclockwise galaxies showed that 10 galaxies classified as clockwise and 13 galaxies classified as counterclockwise did not have identifiable spin patterns, but none of these galaxies was missclassified.

Separating the galaxies to clockwise and counterclockwise galaxies provided a dataset of 87,509 galaxies with clockwise spin patterns and 85,374 galaxies with counterclockwise patterns. The rest of the galaxies were not assigned with identifiable spin pattern, and were not used in the following stages of the experiment. Assuming random 0.5 probability of the galaxy to have each of the two possible spin patterns, the probability to have such separation by chance can be computed using cumulative binomial distribution, such that the number of tests is 172,883 and the probability of success is 0.5. Under these conditions, the two-tailed  probability to have 87,509 or more successes is $P\simeq3.7\cdot10^{-9}$. Repeating the experiment after mirroring the galaxies led to the exact same results. That is expected since Ganalyzer is a deterministic algorithm that works in a fully symmetric manner.

\begin{figure*}
\includegraphics[scale=0.5]{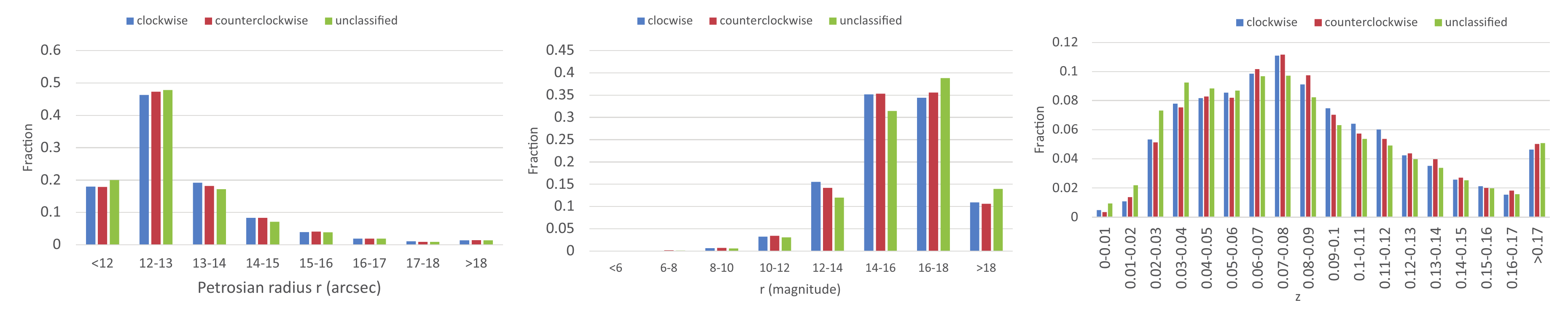}
\caption{Distribution of the r magnitude, Petrosian radius measured in the r band, and the distribution of redshift.}
\label{distribution}
\end{figure*}

Figure~\ref{distribution} shows the distribution of the r magnitude, Petrosian radius measured in the r band, and the redshift of the galaxies classified by Ganalyzer as clockwise, counterclockwise, and galaxies that could not be classified to any of these classes and remained unclassified. The vast majority of the galaxies do not have spectra, and therefore just the subset of 10,281 galaxies that had spectroscopic information could be used for deducing the redshift distribution.

\begin{figure}[h!]
\includegraphics[scale=0.6]{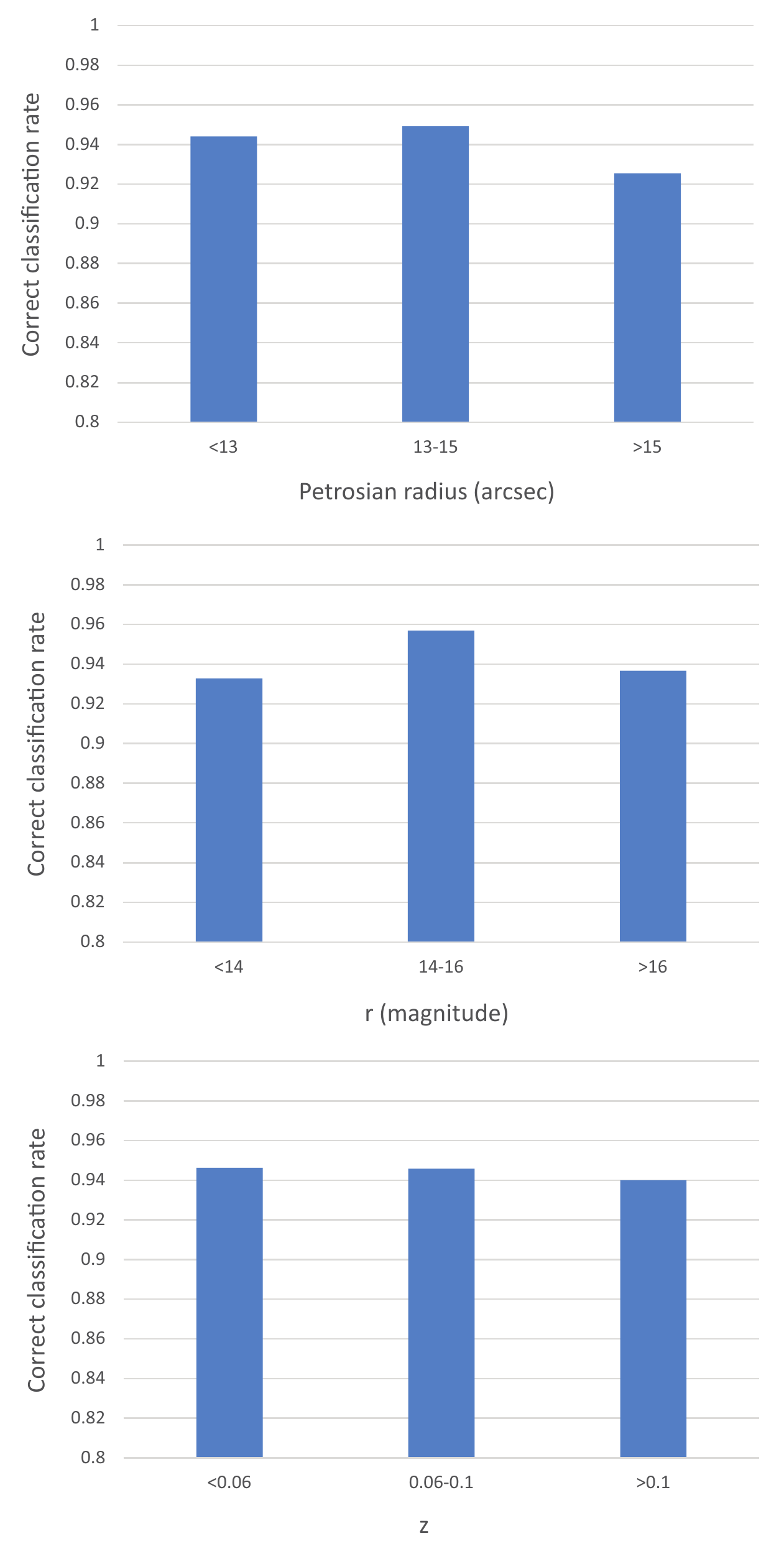}
\caption{Distribution of the galaxies that did not have clear identifiable spin patterns in different r magnitude, Petrosian radius measured in the r band, and redshift.}
\label{dataset_accuracy}
\end{figure}

As the figure shows, while galaxies with higher r magnitude tend to be classified less frequently into clockwise or counterclockwise galaxies, the distribution of the galaxies that could not be classified by {\it Ganalyzer} is largely aligned with the distribution of the galaxies that were classified as clockwise or counterclockwise. Figure~\ref{dataset_accuracy} shows the distribution of galaxies that did not have clear identifiable spin pattern in different redshifts, radii, and r magnitudes.

\section{Results}
\label{results}

As discussed in Section \ref{sec:data}, the difference in the number of clockwise and counterclockwise galaxies is unlikely to be the result of mere chance. Table~\ref{ra_asymmetry} shows the number of clockwise and counterclockwise galaxies in different parts of the sky. The right ascension and declination were separated into 30$^o$ ranges, and the asymmetry measured by $\frac{cw-ccw}{cw+ccw}$ in the different $30^o\times30^o$ sections are specified in the table. The specific parts of the sky in most cases do not show statistically significant differences between the number of clockwise and counterclockwise galaxies, possibly due to the much smaller number of galaxies in each section compared to the entire sky. Table~\ref{ra_numbers} shows the number of galaxies in each of the parts of the sky of Table~\ref{ra_asymmetry}.

\begin{table*}[ht]
\caption{The asymmetry between the number of clockwise and counterclockwise galaxies $\frac{cw-ccw}{cw+ccw}$ in different RA and declination ranges.}
\label{ra_asymmetry}
\begin{center}
{
\scriptsize
\begin{tabular}{lcccc}
\hline
& & Declination (degrees) \\
RA (degrees) & -30-0 &  0-30 &  30-60 &  60-90 \\
 \hline
0-30 & 0.005$\pm$0.01 & -0.015$\pm$0.01 & -0.007$\pm$0.02 & 0$\pm$0.00 \\
30-60 & 0.118$\pm$0.01 & 0.023$\pm$0.01 & 0.025$\pm$0.03 & 0.105$\pm$0.11 \\
60-90 & -0.074$\pm$0.03 & -0.009$\pm$0.02 & 0.149$\pm$0.15 & 0.061$\pm$0.07 \\
90-120 & -0.045$\pm$0.07 & 0.046$\pm$0.02 & -0.011$\pm$0.02 & -0.101$\pm$0.06 \\
120-150 & 0.031$\pm$0.03 & 0.018$\pm$0.01 & 0.028$\pm$0.01 & 0.023$\pm$0.04 \\
150-180 & 0.099$\pm$0.02 & -0.004$\pm$0.01 & -0.008$\pm$0.01 & -0.002$\pm$0.03 \\
180-210 & 0.015$\pm$0.02 & 0.005$\pm$0.01 & -0.001$\pm$0.01 & 0.11$\pm$0.03 \\
210-240 & 0.046$\pm$0.03 & 0.004$\pm$0.01 & 0.046$\pm$0.01 & 0.031$\pm$0.04 \\
240-270 & -0.031$\pm$0.04 & 0.042$\pm$0.01 & -0.028$\pm$0.01 & -0.103$\pm$0.03 \\
270-300 & 0.152$\pm$0.17 & -0.055$\pm$0.08 & 0.114$\pm$0.06 & -0.024$\pm$0.05 \\
300-330 & -0.07$\pm$0.02 & -0.012$\pm$0.01 & -0.107$\pm$0.09 & -0.373$\pm$0.12 \\
330-360 & 0.037$\pm$0.01 & 0.009$\pm$0.01 & -0.047$\pm$0.03 & 0.2$\pm$0.18 \\
\hline
\end{tabular}
}
\end{center}
\end{table*}

\begin{table}[ht]
\caption{The number of galaxies in different RA and declination ranges.}
\label{ra_numbers}
\begin{center}
{
\scriptsize
\begin{tabular}{lcccc}

\hline
& & Declination (degrees) \\
RA ($^o$) & -30-0 &  0-30 &  30-60 &  60-90 \\
 \hline
0-30 & 10482 & 14582 & 1875 & 0 \\
30-60 & 8467 & 8337 & 962 & 76 \\
60-90 & 1425 & 1793 & 47 & 179 \\
90-120 & 224 & 2610 & 2101 & 316 \\
120-150 & 1422 & 10560 & 8063 & 575 \\
150-180 & 2110 & 9523 & 6421 & 1383 \\
180-210 & 1880 & 9271 & 5725 & 1439 \\
210-240 & 1499 & 10968 & 5759 & 512 \\
240-270 & 510 & 6373 & 6176 & 1059 \\
270-300 & 33 & 146 & 271 & 416 \\
300-330 & 3467 & 4636 & 121 & 67 \\
330-360 & 5989 & 11773 & 1230 & 30 \\
\hline
\end{tabular}
}
\end{center}
\end{table}

A cosmological-scale dipole axis is expected to exhibit itself in the form of cosine dependence \citep{longo2011detection,shamir2012handedness}. To test for a possible axis of asymmetry, for each $(\alpha,\delta)$ combination the $\cos(\phi)$ galaxies were fitted into $d\cdot|\cos(\phi)|$, such that $\phi$ is the angular distance between the geocentric coordinates of the galaxy and $(\alpha,\delta)$, and $d$ is the spin direction (1 or -1). That was done by assigning each galaxy with a random number within \{-1,1\}, and $\chi^2$ fitting $d\cdot|\cos(\phi)|$ to $\cos(\phi)$, such that $d$ is the randomly assigned spin direction (1 or -1). The $\chi^2$ was computed 1000 times for each $(\alpha,\delta)$, and the mean and $\sigma$ were computed for each $(\alpha,\delta)$ combination. Then, the $\chi^2$ mean computed with the random spin patterns was compared to the $\chi^2$ when $d$ was assigned to the actual spin direction. The $\sigma$ difference between the mean $\chi^2$ when the spin directions are assigned randomly and the $\chi^2$ determined using the actual spin patterns of the galaxies determines the statistical likelihood of an axis to be at the $(\alpha,\delta)$ coordinates.

\begin{figure}[!ht]
\includegraphics[scale=0.7]{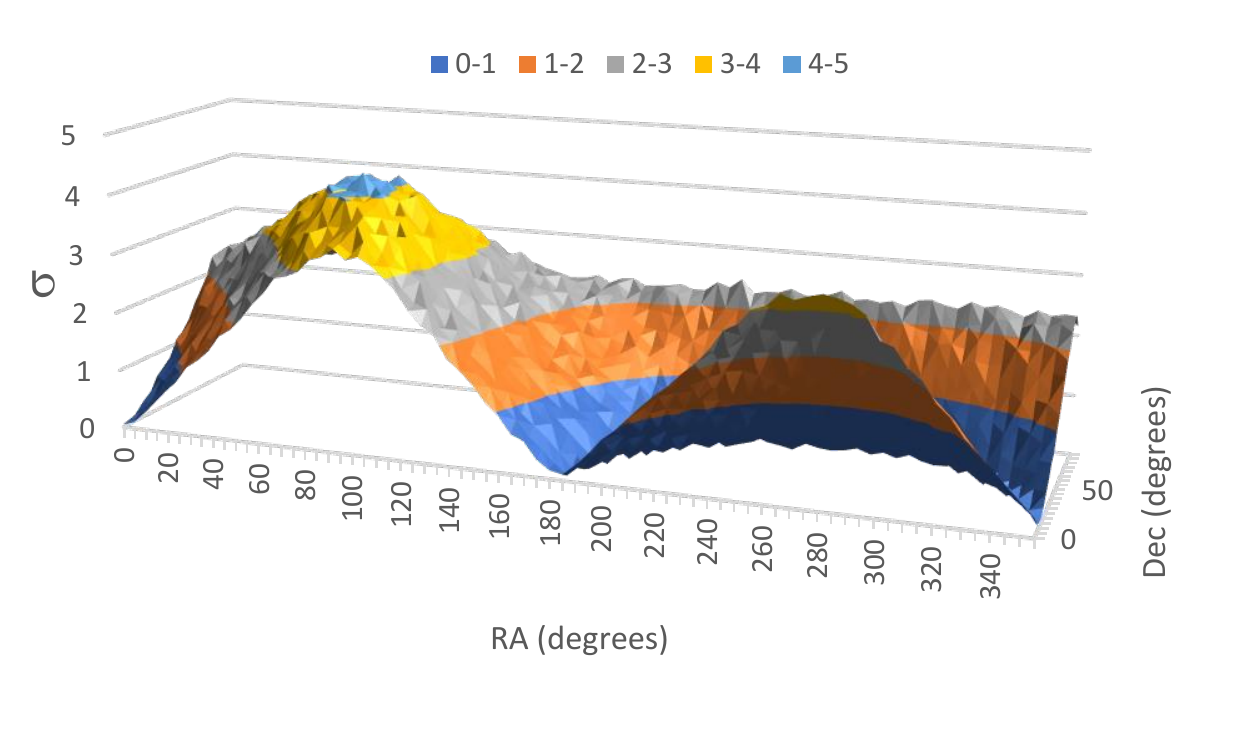}
\caption{The $\sigma$ of possible dipole axes in different $(\alpha,\delta)$ combinations.}
\label{axis}
\end{figure}

Figure~\ref{axis} shows the $\sigma$ of the asymmetry axis of all $(\alpha,\delta)$ combinations. The most likely $(\alpha,\delta)$ was identified at $(\alpha=88^o,\delta=36^o)$, with $\sigma$ of $\sim$4.34 $(P<0.000014)$. 
The 1$\sigma$ error of the axis is $(62^o,124^o)$ for the right ascension, and $(7^o,69^o)$ for the declination. 

\begin{figure}[ht]
\includegraphics[scale=0.7]{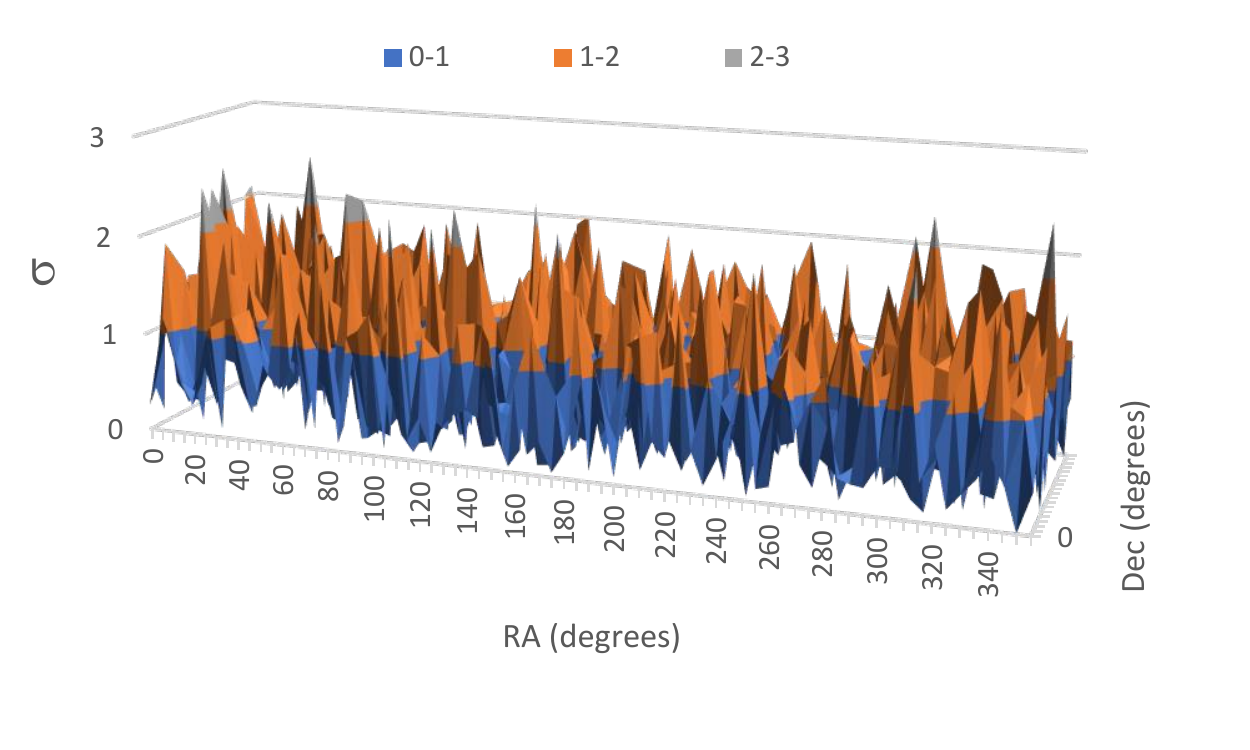}
\caption{The $\sigma$ of the dipole exes in different $(\alpha,\delta)$ combinations such that galaxies were assigned with random spin patterns.}
\label{axis_random}
\end{figure}

Figure~\ref{axis_random} shows the result of the same experiment, but instead of using the spin patterns determined by Ganalyzer as described in Section~\ref{sec:data}, each galaxy was assigned a random spin direction. As expected, the graph does not show any specific pattern or a certain axis that can be considered the most likely axis. The maximum $\sigma$ of all possible $(\alpha,\delta)$ was not higher than 2.5.

\section{Conclusion}
\label{conclusion}

The results of the experiment show asymmetry between the number of galaxies with opposite spin patterns. The portion of the universe observed in this study is far larger than a galaxy supercluster or any other known astrophysical structure, and therefore if the source of the observation is indeed asymmetry between the number of galaxies with opposite spin patterns, that can be considered an evidence of violation of the cosmological isotropy assumption. 

Previous observations using the asymmetry between the number of clockwise and counterclockwise galaxies also showed evidence of violation of isotropy at a cosmological scale \citep{longo2011detection,shamir2012handedness,shamir2013color,hoehn2014characteristics,shamir2016asymmetry,shamir2017colour,shamir2017photometric,shamir2017large,lee2019mysterious,shamir2019photometric,shamir2019photometric2}. The spin pattern of a galaxy is an indication of the galaxy's actual spin direction \citep{iye2019spin}, and therefore asymmetry between spin patterns might indicate on links between the rotation of galaxies that are too far from each other to have gravitational interactions. Cosmological-scale links were observed through other messengers such as gamma ray bursts (GRBs), providing evidence of non-uniform distribution that could violate the isotropy assumption of the cosmological principle \citep{meszaros2019oppositeness}. Short gamma ray bursts (SGRBs) tend to have redshift of $\sim$1 \citep{d2014complete}, but evidence of non-uniform redshift distribution of gamma ray bursts has been observed at redshift of $\sim$2 \citep{horvath2014possible}. Data from the Burst And Transient Source Experiment showed anisotropy in the distribution of SGRBs \citep{meszaros1999existence,balazs1999intrinsic,vavrek2008testing}. On the other hand, long gamma ray bursts (LGRBs) that are typically of higher redshift are distributed more homogeneously \citep{meszaros1995cosmological,kinugawa2019long}.

Fast radio bursts (FRBs) also showed certain evidence of non-homogeneous distribution that might be in violation with the cosmological principle \citep{katz2017frb}. Multiple observations of cosmic microwave background (CMB) data shows possible cosmological-scale polarization \citep{aghanim2014planck,hu1997cmb,cooray2003cosmic,ben2012parity,eriksen2004asymmetries}. 

Violation of the isotropy and homogeneity assumptions can also be related to the conflicting measurements of the rate of the expansion of the universe. Attempts to measure the expansion rate (Hubble constant) provided different results that depend on the messenger, such that measurements with the cosmologically close Cepheids and Tip of the Red Giant Branch (TRGB) provide different results, and these measurements are significantly different from the expansion rate measured using CMB \citep{freedman2019carnegie}. As these measurements are more accurate than in previous years, it is becoming increasingly more difficult to explain the differences in the results without violating the basic cosmological assumptions.

An early attempt to identify patterns in the distribution of galaxy morphology was done by \cite{binggeli1982shape}. The position angles of 44 galaxy clusters from the \cite{abell1958distribution} catalog were determined from elliptical galaxies in the clusters, and the analysis showed that the orientation of galaxy clusters is related to their neighboring clusters to a scale of $\sim$100 Mpc. The average redshift of these galaxies was $\sim$0.071, which is close to the $\sim$0.068 average redshift of the galaxies used in the experiment described in this paper. The non-random pattern of distribution of galaxy spin directions is another indication of a violation of homogeneity in this redshift range, which is clearly far larger than a supercluster or any other known astrophysical structure.

The distribution of the spin directions of the galaxies is aligned in a manner that exhibits a cosmological axis. The most probable dipole axis is identified in $(\alpha=88^o,\delta=36^o)$, with 1$\sigma$ error of $\sim30^o$ in the RA. That distribution of the spin directions can be linked to different geometrical models of the universe \citep{campanelli2006ellipsoidal}, and can also be an indication of previously proposed theories of a rotating universe \citep{godel1949example,ozsvath1962finite,ozsvath2001approaches}. It has been also proposed that asymmetry between spin patterns of galaxies can be driven by parity-breaking gravitational waves \citep{biagetti2020primordial}. While these cosmological models shift from the standard models, direct evidence of the existence of dark matter have not yet been reported, gradually reinforcing the need for models that are not necessarily based on the existence of dark matter.

The results reported in this paper do not have an immediate explanation based on the standard models. However, it is difficult to explain the results by a computer error. Repeating the experiment with randomly assigned spin patterns leads to no identifiable axis. Mirroring the galaxy images leads to flipped numbers of clockwise and counterclockwise galaxies, which is expected as the image analysis algorithm is a deterministic and symmetric model-driven algorithm that is not based on complex rules determined by training data of a machine learning system. Moreover, a computer error is expected to exhibit itself in the form of a consistent bias, rather than different asymmetry in different directions of observation that forms an axis.

The analysis is based on the data acquired by the Sloan Digital Sky Survey, and on the assumption that these data are not biased in some mysterious way that leads to a different number of clockwise and counterclockwise galaxies. It is difficult, however, to think of a bias that would lead to asymmetry between clockwise and counterclockwise galaxies, as none of the SDSS measurements are expected to be sensitive to the spin direction of the galaxy. The galaxy classification is done in a fully automated process, and with no human involvement that could expose the results to human perception bias. More powerful sky surveys such as Vera Rubin Telescope will provide more data, allowing higher resolution profiling of the asymmetry.

\section*{Acknowledgments}

I would like the thank the anonymous reviewers for the comments that helped to improve the manuscript. This study was supported in part by NSF grants AST-1903823 and IIS-1546079. Funding for the Sloan Digital Sky Survey IV has been provided by the Alfred P. Sloan Foundation, the U.S. Department of Energy Office of Science, and the Participating Institutions. SDSS-IV acknowledges support and resources from the Center for High-Performance Computing at the University of Utah. The SDSS web site is www.sdss.org.

SDSS-IV is managed by the Astrophysical Research Consortium for the Participating Institutions of the SDSS Collaboration including the Brazilian Participation Group, the Carnegie Institution for Science, Carnegie Mellon University, the Chilean Participation Group, the French Participation Group, Harvard-Smithsonian Center for Astrophysics, Instituto de Astrof\'isica de Canarias, The Johns Hopkins University, Kavli Institute for the Physics and Mathematics of the Universe (IPMU) / 
University of Tokyo, the Korean Participation Group, Lawrence Berkeley National Laboratory, Leibniz Institut f\"ur Astrophysik Potsdam (AIP), Max-Planck-Institut f\"ur Astronomie (MPIA Heidelberg), Max-Planck-Institut f\"ur Astrophysik (MPA Garching), Max-Planck-Institut f\"ur Extraterrestrische Physik (MPE), National Astronomical Observatories of China, New Mexico State University, New York University, University of Notre Dame, Observat\'ario Nacional / MCTI, The Ohio State University, Pennsylvania State University, Shanghai Astronomical Observatory, United Kingdom Participation Group, Universidad Nacional Aut\'onoma de M\'exico, University of Arizona, University of Colorado Boulder, University of Oxford, University of Portsmouth, University of Utah, University of Virginia, University of Washington, University of Wisconsin, Vanderbilt University, and Yale University.



\bibliography{asym_axis}

\end{document}